\documentclass[aps,prx,twocolumn,floatfix,superscriptaddress,showpacs,showkeys]{revtex4-1}
\usepackage{float,epsf,amsmath,amssymb,verbatim,color,multirow,pifont,graphicx,cleveref,tabularx,cancel,soul}
\usepackage{soul}
\usepackage{color}

\crefname{equation}{Eq.}{Eqs.}
\crefname{section}{Sec.}{Secs.}
\crefname{figure}{Fig.}{Figs.}
\crefname{table}{Table}{Tables.}

\newcommand{\ps}{P_{S}}
\newcommand{\pw}{P_{W}}
\newcommand{\pr}{P_{R}}

\newcommand{\self}{self-consistency }

\begin{document}

\title{Universal mechanism for hybrid percolation transitions}
\author{Deokjae Lee}
\affiliation{CCSS,  CTP and Department of Physics and Astronomy, Seoul National University, Seoul 08826, Korea}
\author{Wonjun Choi}
\affiliation{CCSS,  CTP and  Department of Physics and Astronomy, Seoul National University, Seoul 08826, Korea}
\author{J. Kert\'esz}
\affiliation{Center for Network Science, Central European University, Budapest, Hungary}
\affiliation{Department of Theoretical Physics, Budapest University of Technology and Economics, Budapest, Hungary}
\author{B. Kahng}
\email{bkahng@snu.ac.kr}
\affiliation{CCSS,  CTP and  Department of Physics and Astronomy, Seoul National University, Seoul 08826, Korea}
\date{\today}

\begin{abstract}
Hybrid percolation transitions (HPTs) induced by cascading processes have been observed in diverse complex systems such as $k$-core percolation, breakdown on interdependent networks and cooperative epidemic spreading models. Here we present the microscopic universal mechanism underlying those HPTs. We show that the discontinuity in the order parameter results from two steps: a durable critical branching (CB) and an explosive, supercritical (SC) process, the latter resulting from large loops inevitably present in finite size samples. In a random network of $N$ nodes at the transition the CB process persists for $O(N^{1/3})$ time and the remaining nodes become vulnerable, which are then activated in the short SC process. This crossover mechanism and scaling behavior are universal for different HPT systems. Our result implies that the crossover time $O(N^{1/3})$ is a golden time, during which one needs to take actions to control and prevent the formation of a macroscopic cascade, e.g., a pandemic outbreak.
\end{abstract}

\pacs{89.75.Hc, 64.60.ah, 05.10.-a}

\maketitle

\section{Introduction}
Percolation is a prototypical model of disorder, which is often used to illustrate the emergence (resilience) of a giant cluster as links between individuals are added (deleted) one by one~\cite{perc_review_2,perc_review}.  A giant cluster at a transition point in the mean field limit can be viewed as a critical branching (CB) tree with unit mean number of offspring~\cite{newman,cohen}. The giant cluster of recovered nodes at a transition point of a simple epidemiological model, the so-called susceptible/infective/removed (SIR) model~\cite{newman}, is an instance of such percolating clusters grown in the CB processes. Percolation transition is known as a generic continuous transition~\cite{nuno}.

In a number of systems, however, the situation is more complex: Hybrid Percolation Transitions (HPTs) occur showing features of both second and first-order phase transitions at a transition point~\cite{dodds,pazo,mendes_sync,mukamel}. Examples include the $k$-core percolation~\cite{kcore1,kcore2,kcore_tri,baxter_prx}, and the cascading failure (CF) model on interdependent networks~\cite{buldyrev,son,baxter_mcc,bianconi}. In those systems, as nodes or links are removed one by one above the transition point, the order parameter, the relative size of the giant component decreases continuously, approaches a nonzero value in a scaling manner at the transition point, where it finally collapses to zero: A HPT occurs. Is there a universal mechanism behind this phenomenon? Can it be formulated in terms of branching processes?  Even though these questions are simple and fundamental, there has been no clear answer yet.   

Recently we showed on the example of the CF model that there are two kinds of critical phenomena related to the HPT~\cite{mcc_dj}. One is carried by the behavior of the finite cascades and the other one by the order parameter (the relative size of the giant cluster). We have to distinguish between ``finite" and ``infinite" avalanches (the latter having the size of the giant cluster).
Once an infinite avalanche occurs, the order parameter collapses to zero. Therefore, occurrence of an infinite avalanche is a distinct feature of a HPT, whereas such infinite avalanche is absent for a second-order percolation transition.  Thus we need to investigate what happens in the system while an infinite avalanche proceeds. 

We recall the results of previous studies on $k$-core percolation~\cite{baxter_prx}  and in interdependent networks~\cite{zhou} about the temporal evolution of the giant cluster. The order parameter decreases rapidly in the early time regime, exhibits a plateau for a long time in the intermediate time regime, and decreases rapidly in the late time regime. Moreover, it was found that infinite avalanches proceed in the form of a CB process for a long time, followed by a supercritical process. There has been considerable effort to explain the mechanism leading to this scenario for specific models \cite{zhou, grassberger_nat_phys, grassberger_pre_2016}. In particular, \cite{grassberger_nat_phys, grassberger_pre_2016} pointed out the importance of large loops in the creation of the supercritical process for an epidemic model. However, it has remained uncovered whether there is a universal mechanism, which explains why, how and when such SC processes occur in the late time regime. Here we address these questions and show that  there indeed exists such a universal mechanism, which governs the generally observed crossover behavior in a large class of HPT models.  

In this paper, we  first investigate the mechanism of the crossover behavior from the CB to SC processes using a simple epidemic model  with two-step contagion processes~\cite{janssen} that exhibits a HPT in Sec. II. After explaining the mechanism of HPT on this model, we will show that the same mechanism occurs in other models. In the following sections,  we will consider $k$-core percolation (Sec. III),  the threshold model (Sec. IV),  and  the cascade failure model on interdependent networks (Sec. V). We will show that the mechanism disclosed in Sec. II also work in the other systems in Sections III-V. The final section will be devoted to the summary.   

\section{Two-step contagion model}

We consider the epidemic model introduced in Ref.~\cite{janssen}, which is a generalization of the so-called susceptible (symbolized as $S$)-infected ($I$)-removed ($R$) (SIR) model by adding a weakened state ($W$) between susceptible states. This model is referred to as the SWIR model. Various aspects of the model were studied in Refs.~\cite{swir_kaist_1,hasegawa_1,hasegawa_2,swir_kaist_2,janssen_2}. Besides the usual reactions $S+I \to 2I$ and $I \to R$ of the SIR model we have the additional reactions: $S+I \to W+I$ and $W+I \to 2I$. The reaction rate from $W$ to $I$ is larger than the rate from $S$ to $I$. Specifically, we start the dynamics on Erd\H{o}s-R\'enyi (ER) random graphs of $N$ nodes with all nodes in state $S$ but one node that is in state $I$.  At time step $n$, a node in state $I$ (denoted as ${I}_n$, where subscript represents generation) is selected randomly, the states of all its neighbors are checked one by one. If the state of a neighbor is $S$, then this state  changes either i) to ${I}_{n+1}$ with probability $\kappa$ or ii) to $W$ with probability $\mu$. If the state of a neighbor is $W$, then the state $W$ changes to ${I}_{n+1}$ with probability $\nu$. We repeat the above process for all nodes in state ${I}_n$ and then the state ${I}_n$ changes to $R$ for each associated node. Then all dynamics at time step $n$ are completed and we move to the next time step $n+1$. This dynamics continues until the system reaches an absorbing state in which no more infectious nodes remain in the system. The order parameter $m(\kappa)$ is defined as the fraction of nodes in state $R$. Under the given reaction probabilities, a HPT occurs if the mean degree $z > 2/(\sqrt{\mu^2+4\mu \nu}-\mu)$ and otherwise a continuous transition occurs. This condition is the same as that obtained in Ref.~\cite{bizhani}. The transition point is $\kappa_c=1/z$. The detailed derivations of the transition point and the condition for the HPT are presented in Appendix~\ref{analytic_solution}.

\begin{figure}[t]
\includegraphics[width=1.0\linewidth]{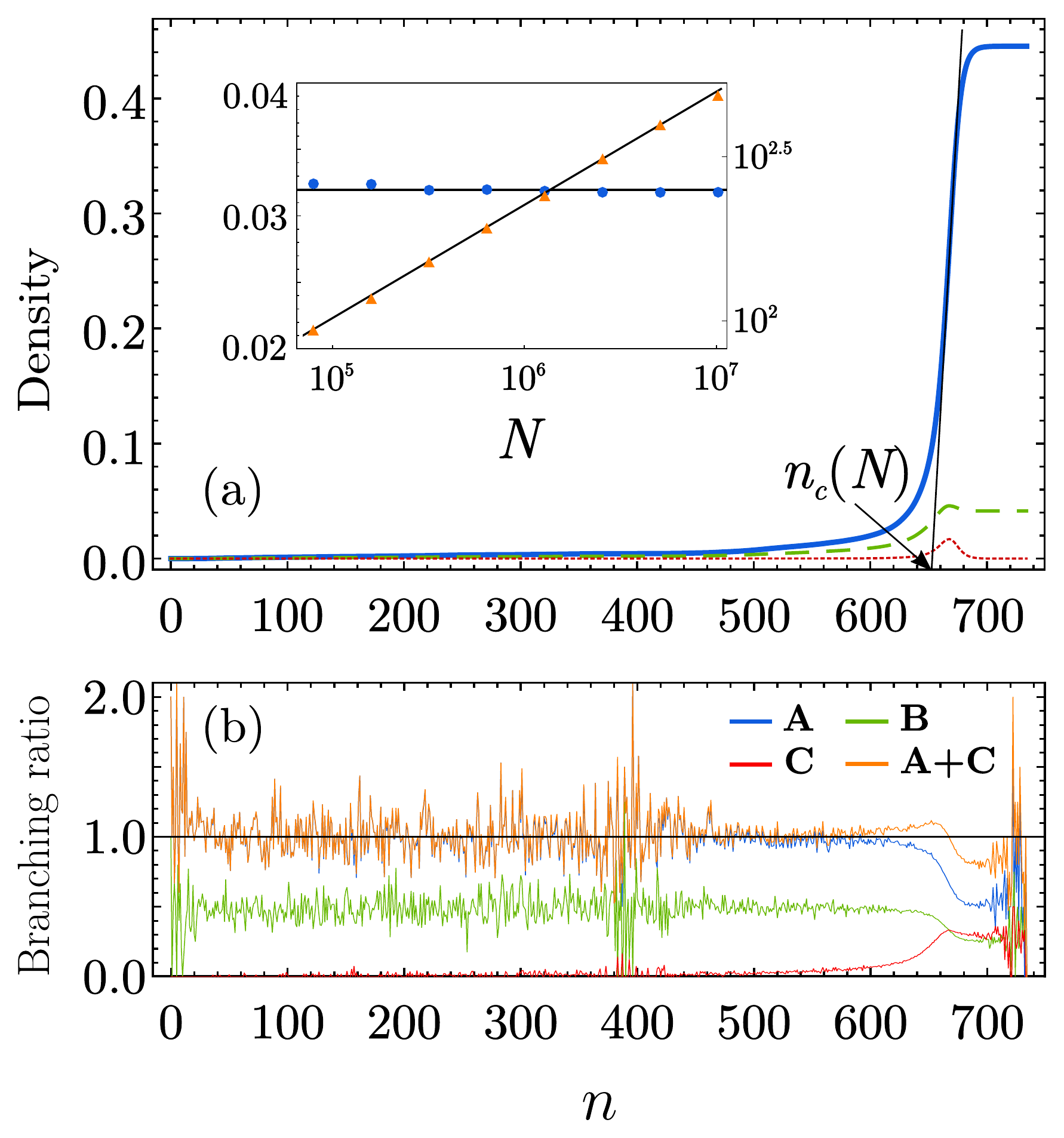}
\caption{(Color online) (a) Plot of the fraction of nodes in states $R$ (blue, solid curve), $W$ (green, dashed curve) and $I$ (red, dotted curve) as a function of generation $n$. Inset: Plot of the maximum slope of the curve ${R}(n)$ vs $N$ ($\bullet$) (left vertical axis). The maximum slopes are independent of $N$. Plot of the characteristic time $n_c(N)$ vs $N$ ($\triangle$) (right vertical axis). 
The fitted straight line has slope 0.35. (b) Plot of the branching ratios as a function of generation $n$ for several types of reactions. Here {\bf A} ({\bf B}) represents the mean number of offspring that change their state from S to $I$ ($W$) by their parents in state $I$. {\bf C} represents the mean number of offspring that change their state from $W$ to $I$ from their parents in state $I$. 
%{\bf A+C} represents the mean number of offspring that change their state from either $S$ or $W$ to $I$ by their parent in state $I$. 
For both (a) and (b), data are obtained from a single realization of an infinite avalanche on an ER network with mean degree $z=8$ of system size $N=5.12\times 10^6$ using the coefficients $\kappa=1/8$, $\mu=1/16$ and $\nu=0.9$.}
\label{fig:branch} 
\end{figure}

\begin{figure}[t]
\includegraphics[width=1.0\linewidth]{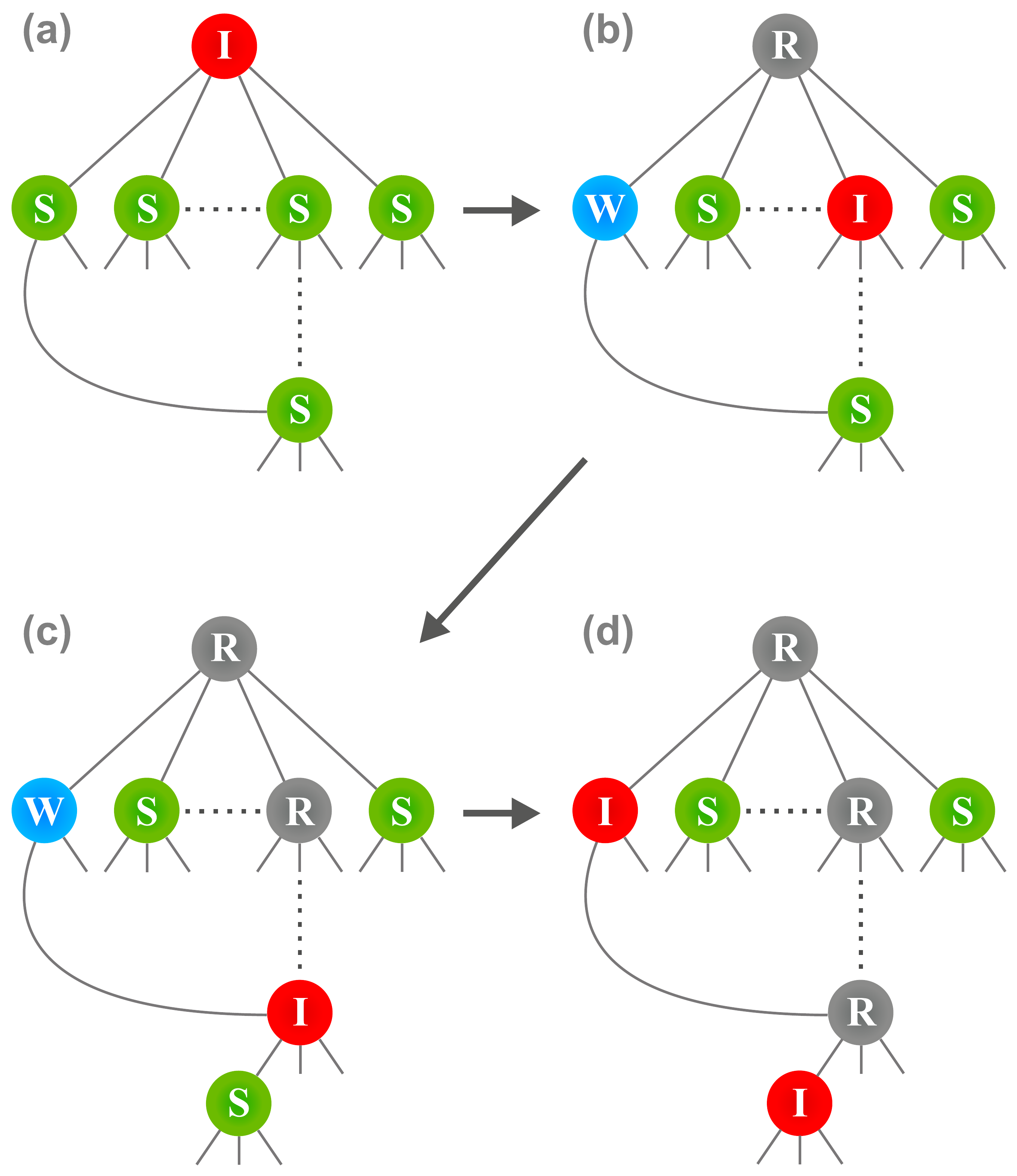}
\caption{(Color online) Schematic picture of the epidemic spreading process of the SWIR  model. (a) 
The process begins from an infectious node. (b) It can infect a susceptible node among its neighbors and change 
the state thereof from $S$ to $I$ and can also change the state of another  neighbor from $S$ to $W$.  This type of process persists for a long time and the critical branching tree is constructed. After a long $O(N^{1/3})$ time passes, an infectious node can contact a node in state $W$ that was created much earlier and change its state from $W$ to $I$ in (d). In addition, (d) the I-node in (c) infects a susceptible neighbor and changes its state to $I$. Thus, a SC process occurs, leading to the jump in the order parameter.}
\label{fig:schematic}
\end{figure}

At the transition point $\kappa_c$, a single infected node can trigger an infinite avalanche of size $O(N)$ with a certain probability $P_{\infty}$. With the remaining probability, finite avalanches occur and their sizes are $o(N)$. When an infinite avalanche occurs, as shown in Fig.~\ref{fig:branch}(a), the order parameter remains almost zero ($o(N)$) for long time up to a characteristic time $n_c(N)$,  beyond which it increases rapidly and reaches its final, $O(N)$ value in a short time period. To see how an infinite avalanche proceeds at a microscopic level, we trace an infection dynamics in the view of branching processes as shown schematically in Fig.~\ref{fig:schematic}. At the step $n=0$, a single infectious seed is present. At each time step, infected and weakened nodes are generated following the aforementioned rule. Because the probability to generate an infectious node per each edge is $1/z$ and the mean number of edges outgoing from the infected parent node is $z$, a single infected node can be generated on average. There is some probability that a weakened node is created. Thus a CB tree of recovered nodes is generated. We notice that, although during the CB process many $W$ nodes are created, there are very few nodes produced from them in state $I$ (consequently nodes in state $R$) through the reaction  ${W+I}\to {I+I}$ 
as shown in Fig.~\ref{fig:branch}(b). However, as the dynamics proceeds and approaches $n_c(N)$, the reaction ${W+I}\to {I+I}$ occurs more frequently and the branching ratio to generate a node in state $I$ through this reaction becomes non-negligible. 

\begin{figure}
\centering
\includegraphics[width=1.00\linewidth]{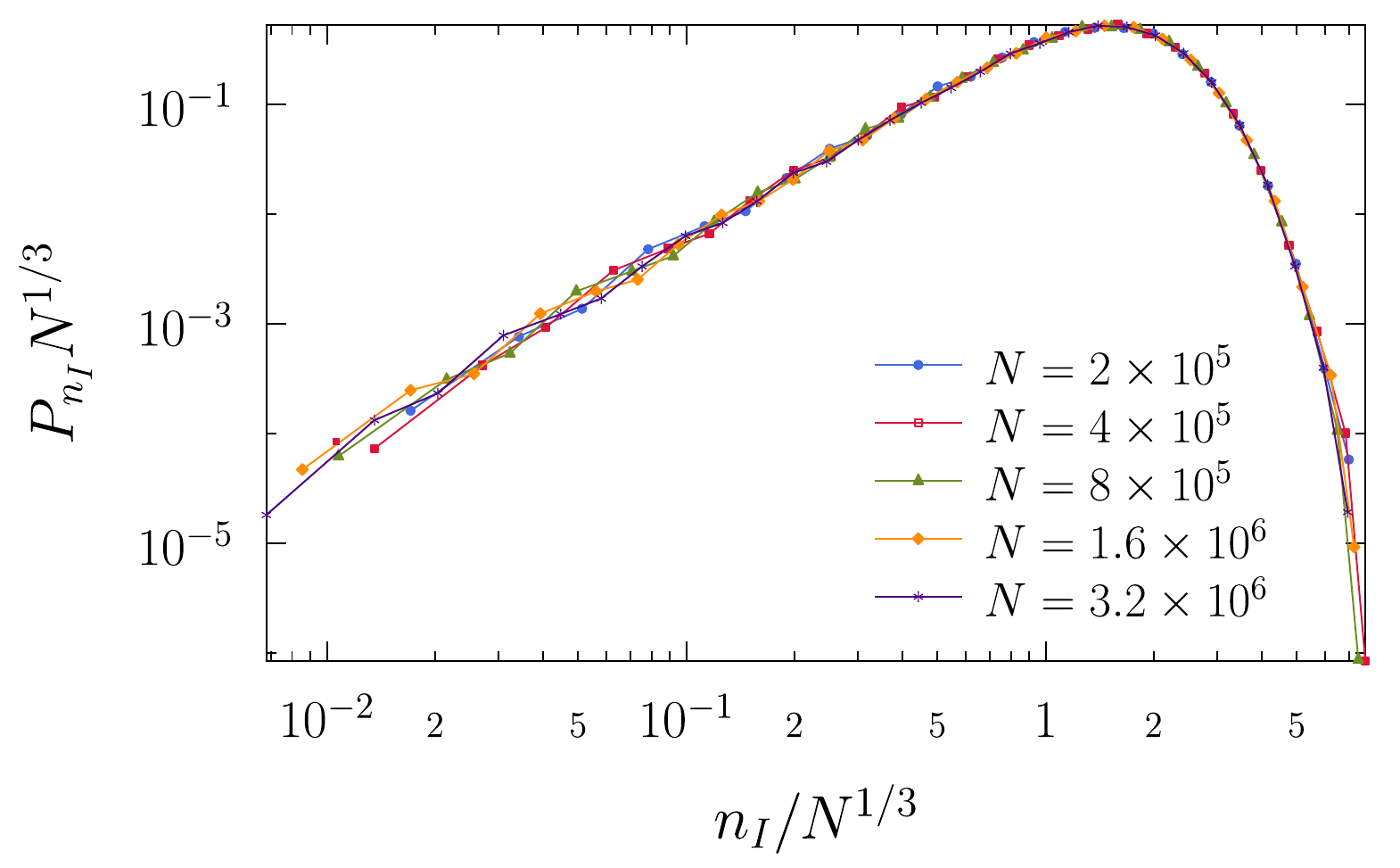}
\caption{(Color online) For the SWIR model, scaling plot of the probability $P_{n_I}$ of the generation $n_I$ at which a loop is formed in critical branching processes on ER networks. Data for different system sizes are well collapsed onto a single curve with the scaling form of $P_{n_I}N^{1/3}$ as a function of $n_I/N^{1/3}$. Data are obtained from ER network with mean degree $z=8$ far away from the transition point $z_c=1$.}
\label{fig:loop}
\end{figure}

In order to determine the crossover point between CB and SC we recall that the size of the largest cluster at the critical average degree $z_c$ of ordinary percolation of the ER graph is $O(N^{2/3})$~\cite{bollobas,dslee}. The giant cluster at criticality has the topology of a CB tree such that the branching process persists up to steps $O(N^{1/3})$ beyond which finite-size effects appear in the form of short-range and long-range loops~\cite{bollobas} (see also Appendices~\ref{cb_size_lifetime} and \ref{app:loop_length}). In the epidemic models on ER networks the average degree is above the percolation threshold ($z>z_c$, otherwise global spreading would be trivially impossible), however, the reaction probability $\kappa_c = 1/z$ assures just the critical branching probability by which the infection proceeds. Thus the growing cluster of $R$ nodes can be considered as if a critical ER cluster would develop on the ER supercritical graph.  In accordance with this picture the probability distribution of the generation at which a loop is formed in CB processes shows a peak at a characteristic generation $n_c(N)\sim O(N^{1/3})$ (Fig.\ref{fig:loop}). This means that long-range loops begin to form mostly when a CB tree is grown up to $n_c(N)$. Based on this, we conclude that before $n_c $ the $I$-state nodes are almost entirely generated through the CB tree and the $W$-state nodes accumulate to an extent of $O(N^{2/3})$ because the number of $W$-state nodes is proportional to that of $I$-state nodes. 
Around $n_c(N)$ the loops become important, and due to the long range links the reaction ${W+I}\to {I+I}$ occurs over the entire system, with $W$-state nodes having been generated at all times to $O(N^{1/3})$ (see Appendix~\ref{app:loop_length}). The accumulated population of $W$-state nodes and the possibility of long-range loop formation lead to an increase of the number of infected offspring above the critical value resulting in the SC process and eventually in the jump of the order parameter. 
We remark that up to the characteristic generation $n_c(N)$, the population of recovered nodes in state $R$ is less than or equal to $O(N^{2/3})$, sublinear to the system size $O(N)$, whereas beyond $n_c(N)$, the population suddenly increases to $O(N)$. Thus, we regard the characteristic generation $n_c(N)$ as {\it the so-called golden time}, during which one needs to take some actions to control and prevent a pandemic outbreak. We also remark that the formation of long-range loops needed for a discontinuous percolation transition was first observed and conjectured in a model of two interacting epidemics~\cite{grassberger_nat_phys,grassberger_pre_2016}. However, the connection between the length scale of long loops and finite-size scaling of the ordinary percolation was missing, so that the scale of golden time could not be predicted.

\begin{figure}[h]
\includegraphics[width=1.0\linewidth]{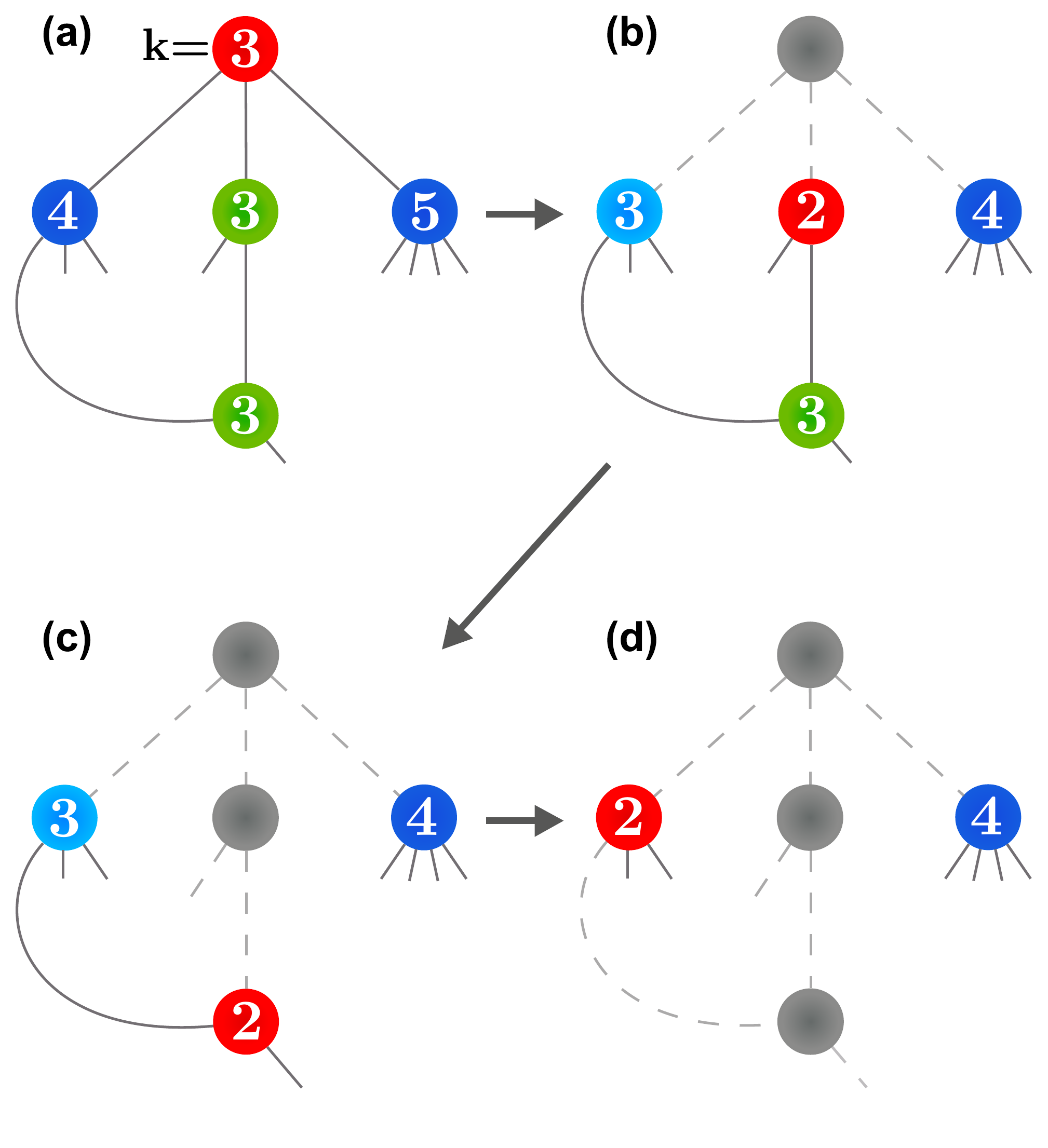}
\caption{Schematic illustration of an avalanche processes of ($k=3$)-core percolation. At the beginning, if degree of a node is $k=3$ ($k>3$), the node is represented by green (dark blue) circle, which corresponds to a susceptible (generalized weakened) node for the SWIR model.  If degree of a node becomes $k=3$ during the dynamics, that node is regarded as a weakened node (light blue). The number in each circle represents the degree of that node at respective step.}
\label{fig:schematic_kcore}
\end{figure}

\section{$k$-core percolation}

$k$-core percolation is known as a prototypical model that exhibits a HPT. The $k$-core subgraph is identified on a graph (here the ER graph with mean degree $z$) as follows. One starts with removing all nodes that have degree less than $k$. These removals may decrease the degrees of remaining nodes. If degrees of some nodes become less than $k$, then those nodes are removed as well. This process is repeated until no more node is removed. For $z > z_c$, a $k$-core subgraph remains after the pruning process and its size is $O(N)$. The order parameter is the relative size of the $k$-core subgraph.

Here we remove a randomly chosen node from the $k$-core subgraph and repeat the pruning process once again. Near $ z_c $, this process can remove all nodes (infinite avalanche of size $O(N)$) or a fraction of nodes (finite avalanche of size $o(N)$) from the $k$-core subgraph, each of which contributes to a discontinuous or continuous change of the order parameter in the thermodynamic limit, leading to a first-order or second-order  transition, respectively~\cite{kcore_dj}.

We focus on the infinite avalanches at $z_c$ from the perspective of branching processes.  Let us consider a $k$-core subgraph configuration at $z_c$  in which each node has at least $k$ degree and the deletion of node $i$ leads to the collapse of the entire system. The node $i$ is regarded as an infectious seed node ($I$). We check the degrees of neighbors of the node $i$. If a neighbor of $i$ has degree $k$, it is regarded as a susceptible node ($S$), and changes its state to $I$ because it will be deleted after node $i$ gets deleted. If a neighbor has degree $\ell > k$, then it is regarded as a generalized weakened node and denoted as ${W}_{\ell-k}$. Now its state changes to $W_{l - k - 1}$. The subscript $\ell-k$ refers to the threshold and decreases as the neighbors of that node are deleted. When it becomes zero during an avalanche, the state ${W}_{\ell-k}$ becomes $W$ and the node has the same role as weakend nodes in the SWIR model. This node gets infected when it contacts an $I$-state node once more.  In analogy with the process in the SWIR model the infective state in the $n$-th generation or branching step is denoted by ${I}_n$. Once the dynamics in $n$-th step is completed, the nodes in state ${I}_n$ are deleted. 

The avalanche processes of ($k$=3)-core percolation is schematically illustrated in Fig.~\ref{fig:schematic_kcore}. At the beginning, all nodes are classified into the two types, susceptible nodes with degree $k=3$ (green circles) and generalized weakened nodes with $k > 3$ (dark blue circles). The dynamics proceeds in the following way: In (a), an infinite avalanche process begins at a node with any degree $k \ge 3$ (red circle). Here we take the case $k=3$. This node corresponds to an infectious seed for the SWIR model. In (b), when that triggering node is intentionally removed (black circle), the degrees of its neighbors are decreased by one. If the degree of a neighbor becomes $k<3$ (red), $k=3$ (light blue), and $k>3$ (dark blue), the node is regarded as an infected node (corresponding to $S+I\to 2I$) and is removed next step (red), is regarded as a weakened (light blue), and remains as a generalized weakened node (dark blue), respectively. In (c), the red node is removed and becomes black. The degree of its  neighbor is again decreased by one. If a degree becomes $k<3$ (red), the susceptible neighbor becomes infected, which is to be removed next step. In (d), the red node in (c) is removed and the degrees of its neighbors are decreased by one.  Here the red node, whose degree was originally four, now becomes infected by the contacts with previously two red nodes. Such reaction processes correspond to $S+I\to W+I$ and $W+I\to 2I$. This removal process is possible only when a loop is formed between the light-blue node and the red node in (c). Loop length is shown rather short in this schematic figure but loops can be long as much as $O(N^{1/3})$ in an ER graph. The processes (a)-(d) continue until no more red node remains.

\begin{figure}[t]
\includegraphics[width=1.0\linewidth]{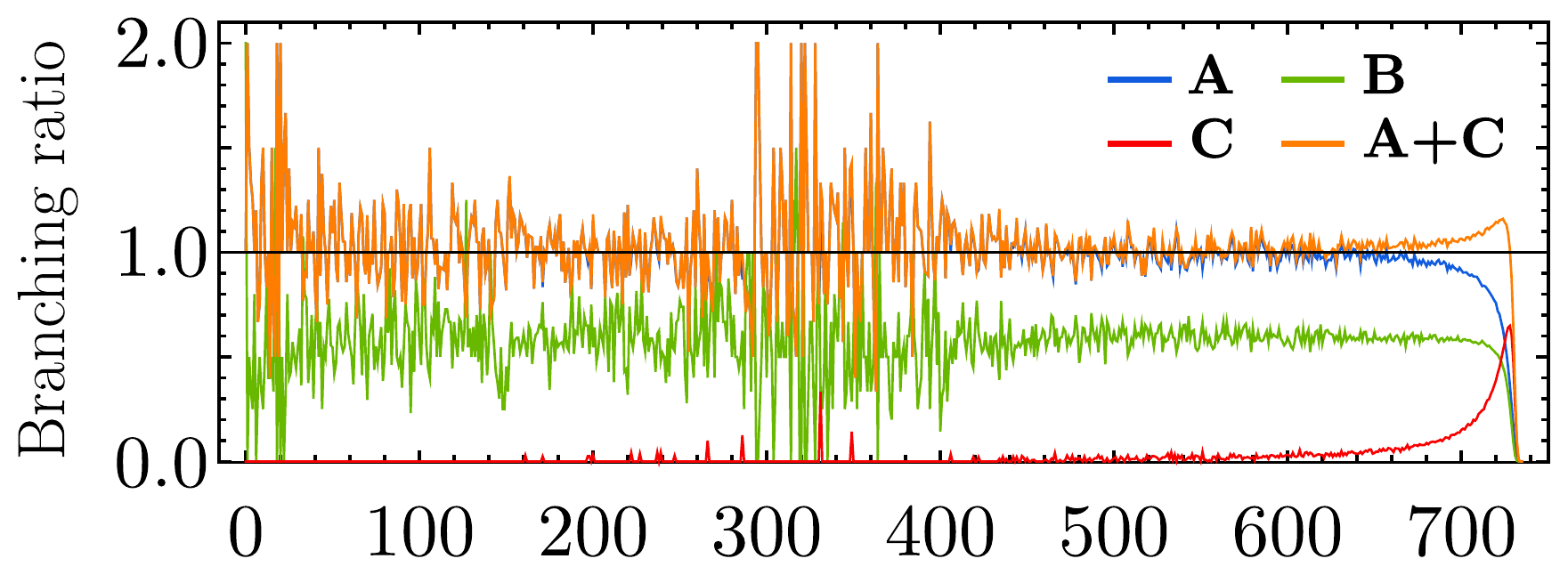}
\caption{(Color online) 
For $k$-core percolation, plot of the branching ratios as a function of  generation $n$ for each type of reactions during an infinite avalanche. {\bf A} represents the ratio of removed nodes with original degree $z=3$ ($z > 3$), which corresponds to the reaction $S+I\to 2I$ for the SWIR model. {\bf B} does the ratio of the reaction $W_{k-l>0}+I \to W+I$, which corresponds to the ratio of generating weakened nodes denoted by $W$. {\bf C} does the ratio of nodes changing their degrees to $k=3$ $(W+I\to 2I)$. $\bf{A + C}$ represents the total branching ratio of $I$. Data are obtained from a network with $N=5.12 \times 10^6$ at a transition point.
}
\label{fig:k_core}
\end{figure}

Fig.~\ref{fig:k_core} shows the branching ratio as a function of branching step $n$ for an infinite avalanche of $k$-core percolation. We find again that the CB  process continues up to a characteristic step $n_c(N)$ when it changes to the SC  process. By the crossover time $n_c$ large number of nodes get their degrees reduced to $k$ so that they become $W$-nodes. The SC  process is again driven by the meeting of an old $W$ node with a new $I$ node, ${W+I}\to {I+I}$. Such a reaction sets up the rapid SC process and the entire collapse of the $k$-core subgraph. 

\section{The threshold model} 

The threshold model was introduced~\cite{watts} for understanding the spread of fads, cultural traits, the diffusion of norms, and innovations, on social networks. In this model, each node $i$ is assigned its threshold value $q_i$ and exists in one of two states, either active or inactive state. An inactive node $i$ with $m_i$ active neighbors among $k_i$ total neighbors (degree) becomes active when its fraction of active neighbors, $m_i/k_i$ exceeds its threshold value $q_i$. This threshold model is known to exhibit a hybrid phase transition when mean degree $z$ is sufficiently large. Here we show that the mechanism underlying this hybrid phase transition is the same as we observed in the previous instances.

To illustrate how the universal mechanism works in the threshold model, we reconsider the rule of the threshold model in the perspective of the SWIR model in the following way: We match up active nodes in the threshold model with either infectious $I$ or recovered $R$ nodes in the SWIR model.  
Among the active nodes, an $I$-state node is the node that becomes active at the preceding step. The other active nodes are regarded as $R$-state nodes. Inactive nodes are matched up with either susceptible $S$ or weakened $W$ nodes in the SWIR model:  i) A node satisfying $k_i q_i <1$ from the beginning is regarded as susceptible node.  ii) A node satisfying $k_i q_i > 1$  is regarded as a generalized weakened node and denoted as $W_{\lfloor k_iq_i\rfloor - m_i}$ similarly to the $k$-core percolation case. Then the dynamics proceeds following the same way as in the $k$-core percolation. 

We performed simulations with a single threshold value $m_i=0.16$ for all nodes on ER networks with mean degree $z_c=7.47707$ which is the transition point for the given threshold value. At this point, the cascade dynamics becomes critical, so that the avalanche size distribution follows a power law. 

\begin{figure}[h]
\includegraphics[width=1.0\linewidth]{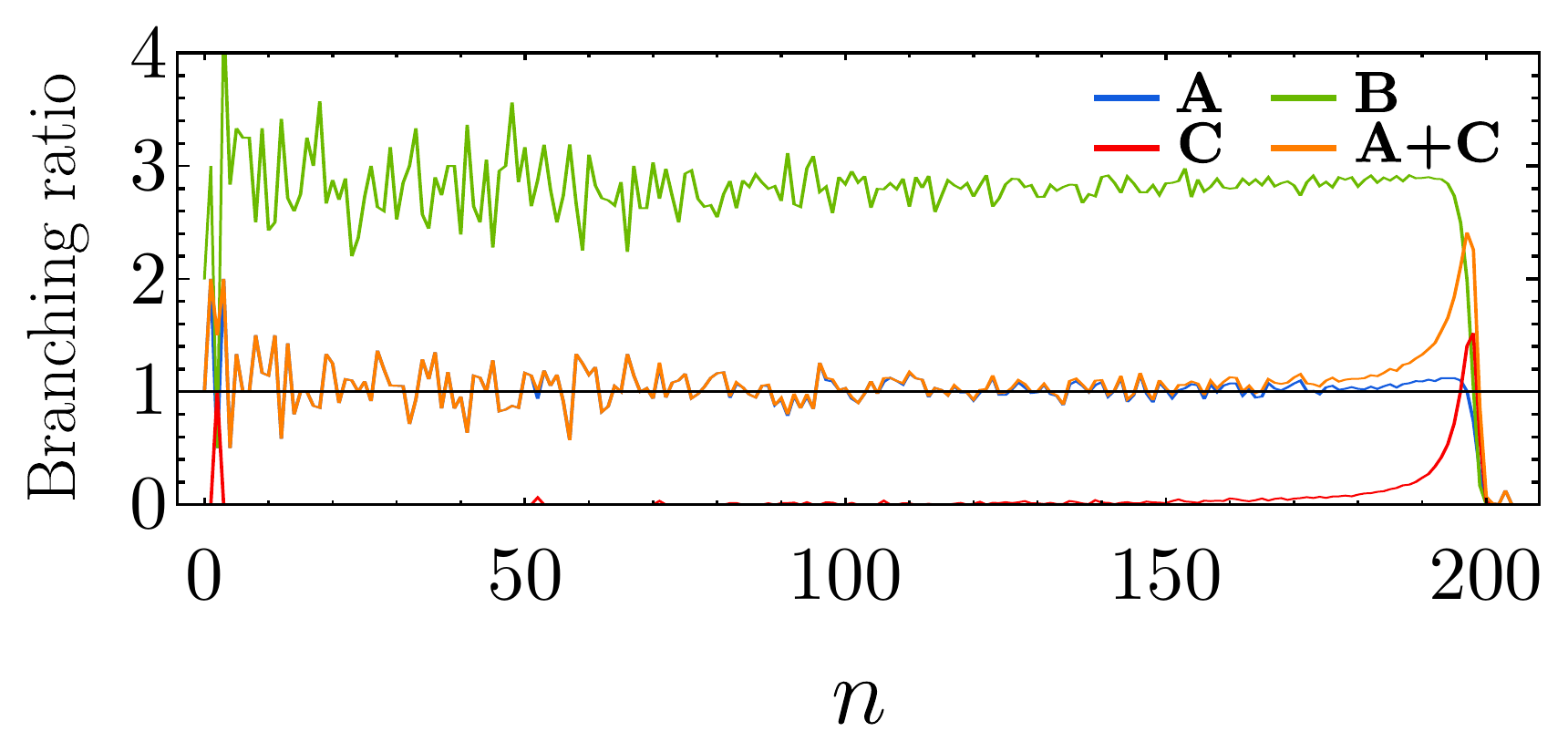}
\caption{For the threshold model, plot of the mean number of offspring as a function of generation $n$ for several types of node deletions. {\bf A} and {\bf B} represent the mean number of $I$ and $W$ offsprings, respectively, generated by the reactions $S+I\to 2I$ and $W_{\lfloor k_iq_i\rfloor - m_i}+I\to W+I$, respectively in terms of the SWIR model. {\bf C} represents the mean number of $I$ offspring generated by the reaction corresponding to $W+I\to 2I$. The sum of the mean $I$ offspring from {\bf A} and {\bf C} exhibits critical branching behavior up to a characteristic generation $n_c(N)$, after which it increases drastically and shows supercritical behavior and then collapses to zero. Thus, the threshold model evolves following the universal mechanism we concern.   
}
\label{fig:threshold}
\end{figure}

We obtain the branching ratios as a function of dynamic step (generation) $n$ for several types of reactions for the threshold model, which is shown in Fig.~\ref{fig:threshold}.
Here we also find a crossover from a CB to SC process similar to that of the SWIR model. Again the accumulation of a sufficient number of weakened nodes during the CB process and their activations through long-range loops are the underlying mechanism of the SC behavior.

\section{The CF model on interdependent networks}

\begin{figure*}[h]
\includegraphics[width=0.8\linewidth]{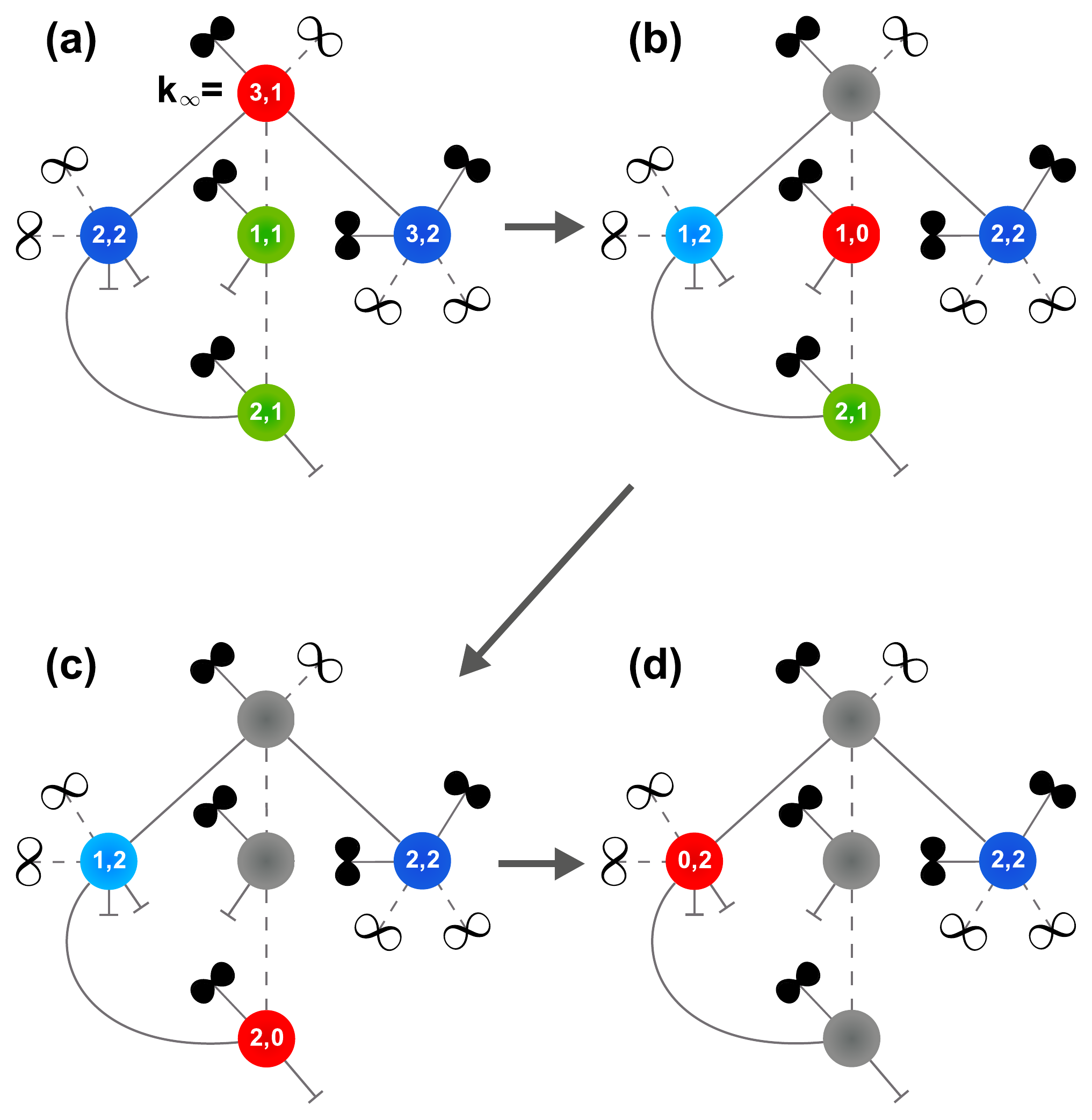}
\caption{For the CF model, schematic illustration of an infinite avalanche process. Nodes with degree one for any type of links are colored by green, which correspond to susceptible nodes. Nodes with degree more than one for any type of links are colored by dark blue, which corresponds to generalized weakened state in $k$-core percolation. The cascading dynamics starts by removing a randomly selected node in any state in (a), denoted by red circle, which is a seed. Then, the effective degrees of the neighbors of the red node are changed as follows: the degree (2,2) change to (1,2),  (1,1) to (1,0) and (3,2) to (2,2). After that, the red node becomes recovered (denoted as black circle). 
The node with updated degrees as (1,2) in (b) changes its color from dark to light blue, representing that the node now becomes weakened, because it can be infected by contacting one more infected node via the solid line (i.e., losing the green neighbor). The node updated as (1,0) no longer belongs to the GMCC and must be removed next step shown in (c). 
The red node in (b) is removed. Consequently, the node with effective degree (2,1) in (b) changes to the degree (2,0) and shall be removed next step. 
The red node in (c) is removed. One of its neighbors with degree (1,2) in (c) changes its degree to (0,2), which is to be removed next step in (d). This corresponds to the reaction $W+I \rightarrow 2I$. This reaction is possible through the long-range loop in (c).  
 }
\label{fig:schematic_mcc}
\end{figure*}

We consider here ER interdependent networks in the single layer representation of Ref.~\cite{son2}. In this picture we have a single ER graph but with two types of links (A and B), for each having the average degree $z$. The order parameter is the relative size of the giant mutually connected cluster (GMCC), in which every pair of nodes are connected following each type of links. The CF model exhibits a HPT at the transition point $z_c$~\cite{baxter_mcc,mcc_dj}.

As with $k$-core percolation, the removal of a node from the GMCC can induce further removal of nodes from the GMCC. This avalanche can be infinite or finite, each of which contributes to the discontinuity of the order parameter or the critical behavior of the HPT, respectively~\cite{mcc_dj}. Here we focus on the infinite avalanches at $z_c$ which leads to the collapse of the entire GMCC.

We consider the avalanche process in the view of a branching process of removed nodes~\cite{baxter_mcc}. To describe the avalanche process in terms of the SWIR model, we determine the effective degrees $k_A(j)$ and $k_B(j)$ of each node $j$ for each type of links. The effective degree $k_A(j)$ ($k_B(j)$) is defined as the number of $A$-type ($B$-type) of links of the node $j$ following which one can reach $O(N)$ nodes. Each node in the GMCC has $k_A \ge 1$ and $k_B \ge 1$. We explain how to determine the effective degrees of each node in simulations in Appendix~\ref{effective_degree}

Specifically, as shown in Fig.~\ref{fig:schematic_mcc},  there exist two types of links represented by solid and dashed lines. Infinity mark (filled or empty $\infty$) on a link (solid or dashed) represents that one can reach $O(N)$ number of nodes through the links of that type. Links with bar ($|$) mean that one can reach only $o(N)$ nodes. Pair of numbers inside the circle of each node represent effective degrees of respective node. Those are the number of solid links and dashed links in order that lead to $O(N)$ number of nodes. If a node loses all such links of any type, it is separated from the GMCC. For instance, the node with the effective degree (1,1) at the middle of Fig.\ref{fig:schematic_mcc}(a) is a member of the GMCC  through its only solid link and the dashed link connected to the red node. 

The cascading dynamics proceeds in the following way: An avalanche is initiated by removing a node chosen randomly from the GMCC. During an avalanche, we identify removed nodes at each time step, then the effective degrees of the neighbors may decrease. As a result one or both type of the effective degrees of some neighbors can become zero. Then, they are removed from the GMCC  at the next time step, i.e., they are infected and removed at the next time step.  Such avalanche process propagates to all neighbors of those infected nodes recursively until no more node is removed. 

If a node is removed at a time step, it is regarded as an $I$-state node and it becomes $R$-state node at the next time step. If one or both effective degrees of a node is unity from the beginning, the node is regarded as a $S$-state node because it can be infected (i.e., removed) by contacting an infected node (i.e., losing the unit effective degree). If one or both effective degrees of a node become unity during an avalanche, we identify the state of that node as $W$ because the node became vulnerable as a result of contacting infected nodes. This view enables us to understand the correspondence between the cascading dynamics of the CF model and the dynamics of the SWIR model. A specific example of the avalanche dynamics is presented in the caption of Fig.~\ref{fig:schematic_mcc}. 

\begin{figure}[t]
\includegraphics[width=1.0\linewidth]{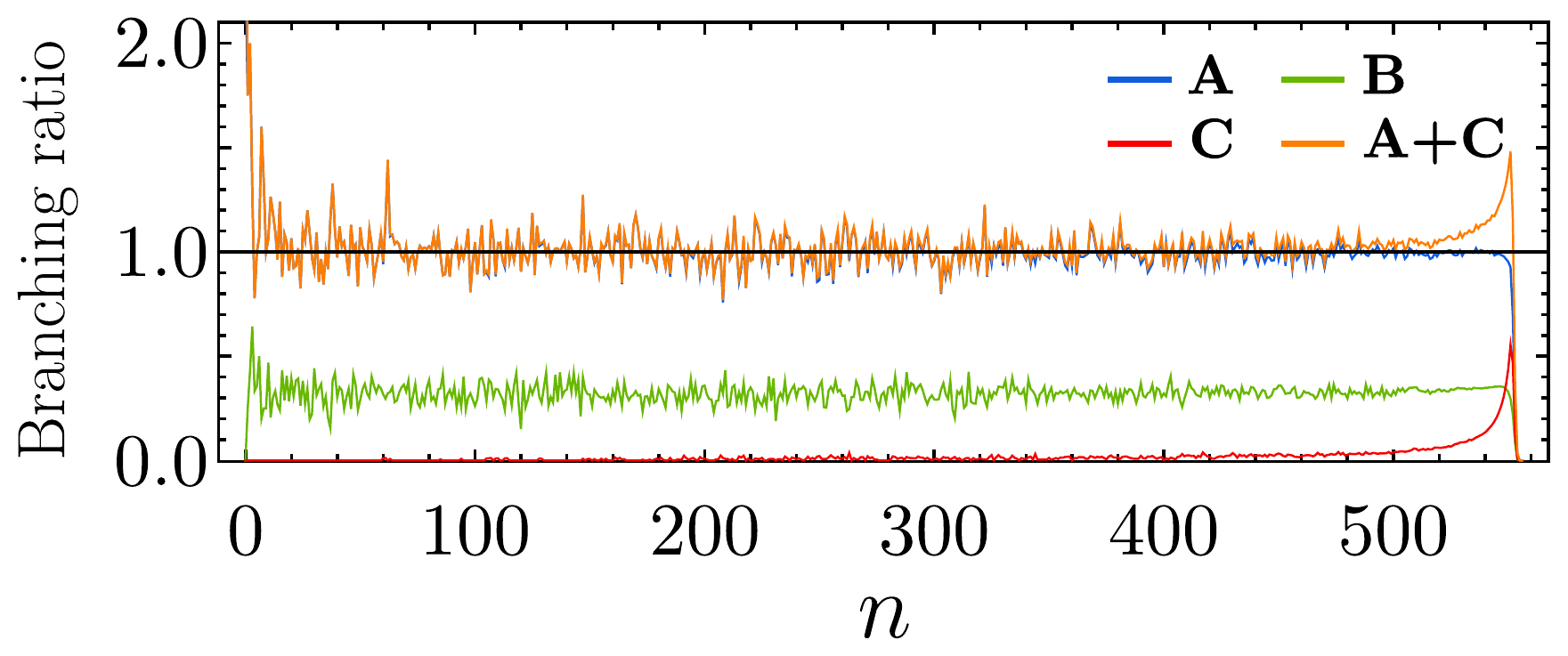}
\caption{(Color online) For the CF model, plot of the branching ratios as a function of  generation $n$ for each type of reactions during an infinite avalanche. {\bf A} represents the mean number of new $I$-state nodes transformed from $S$-state nodes ($S + I \to 2I$). {\bf B} is the mean number of new $W$-state nodes ($k_A$ or $k_B$ becomes unity for the first time). {\bf C} is the mean number of $I$-state nodes transformed from $W$ nodes ($W + I \to 2I$). $\bf{A + C}$ represents the total branching ratio of $I$. Each figure is obtained from a network with $N=5.12 \times 10^6$ at a transition point.
}
\label{fig:branching_mcc}
\end{figure}

Fig.~\ref{fig:branching_mcc} shows the branching ratio as a function of branching steps $n$ for an infinite avalanche of the CF model. We find that a CB process persists and the generating ratio of the weakened nodes is constant with some fluctuations. The number of infected offsprings from weakened nodes is negligible up to a characteristic step $n_c(N)$, beyond which it increases rapidly. Thus the generation ratio of infected offspring exceeds unity beyond $n_c(N)$: a collapse of the giant MCC takes place. 

\begin{figure}[h]
\includegraphics[width=1.0\linewidth]{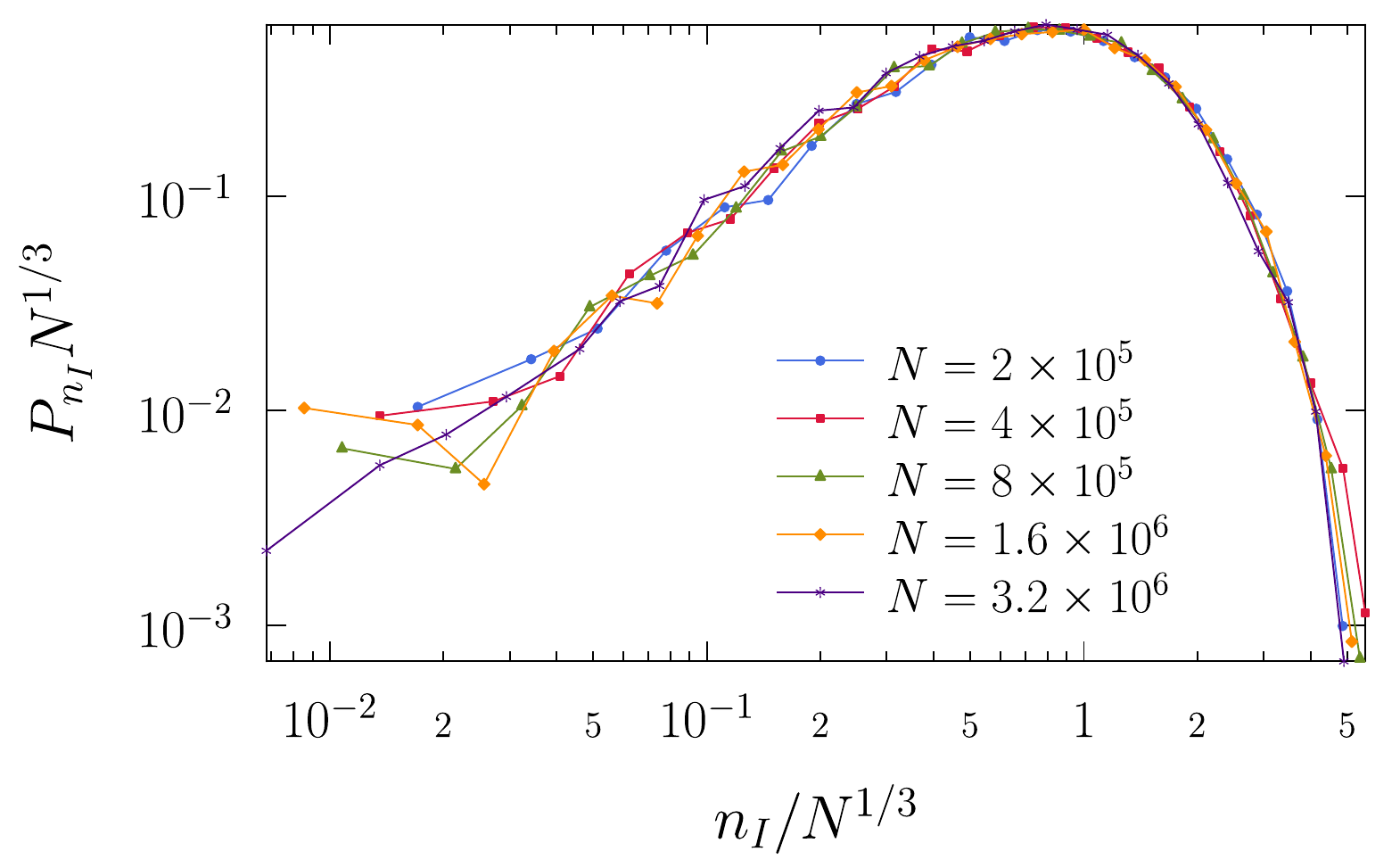}
\caption{
For the CF model, scaling plot of the probability distribution $P_{n_I}$ of the generation $n_I$ at which a node in state $W$ becomes $I$ for the finite avalanche case. Data for different system sizes are well collapsed onto a single curve with the scaling form of $P_{n_I}N^{1/3}$ as a function of $n_I/N^{1/3}$. This means that the loops of length $O(N^{1/3})$ are abundant.}
\label{fig:loop_mcc}
\end{figure}

In order to determine the crossover point between CB and SC for the cascade dynamics in the CF model,  we examine the probability distribution that a node in state $W$ at generation $n$ changes to its state to $I$ in Fig.~\ref{fig:loop_mcc}. We find that the change mainly occurs at the generation $O(N^{1/3})$ in finite avalanches. This is actually expected because finite size effect arises at the generation $O(N^{1/3})$

\section{Summary}
We disclosed the universal mechanism of the HPT induced by cascade dynamics on ER networks. We have shown that during the CB processes, the order parameter sustains up to the time step of $O(N^{1/3})$, and $W$-state nodes accumulate to an extent of $O(N^{2/3})$. After the CB processes, those $W$-state nodes change their state to state $I$ in the way of a SC process. Such reactions are achieved along long-range loops of length $O(N^{1/3})$ presented in finite systems. As a consequence, infected nodes are generated abundantly in a short time, leading to a discontinuity of the order parameter. This explains that the SWIR model and $k$-core percolation do not exhibit discontinuous transitions in low dimensional Euclidean space because of the absence of long-range connections in such space. Next, we showed that the mechanism is universal for diverse systems such as the multi-stage contagion models including the SWIR model and the threshold model, $k$-core percolation and the CF model on the interdependent networks. We expect more models also to belong to this category.  Finally, we regarded the characteristic generation $n_c(N)\sim O(N^{1/3})$ as {\sl the golden time} during which one can control a pandemic outbreak of macroscopic disaster, because for $n < n_c(N)$, the number of damaged nodes is sublinear as $O(N^{2/3})$  to the system $O(N)$, while for $n > n_c(N)$,  it is linear as $O(N)$.

\begin{acknowledgments}
This work was supported by the National Research Foundation of Korea by grant no. NRF-2014R1A3A2069005. JK thanks APCTP and SNU for kind hospitality as well as H2020 FETPROACT-GSS CIMPLEX Grant No. 641191 for partial support.
\end{acknowledgments}

\appendix

\section{Analytic results for the HPT of the SWIR model}\label{analytic_solution}

\subsection{Transition point}

In an absorbing state, each node is in one of three states, the susceptible S,  weakened W and recovered R states. We consider the probability $\ps(\ell)$ that a randomly selected node is in state S when it contacts $\ell$ neighbors in state R.  This probability means that the node remains in state S even though it has been in contact $\ell$ times with those $\ell$ neighbors in state I before they change their states to R. Thus we obtain
\begin{equation}
\ps(\ell) = (1-\kappa-\mu)^{\ell},
\label{p_s}
\end{equation}
where $ \kappa $($ \mu $) is the reaction probability of S becoming I(W) by single attack. Next, $P_{\rm W}(\ell)$ is similarly defined as the probability that a randomly selected node is in state W after it contacts $\ell$ neighbors in state I before they change their states to R. The probability $\pw(\ell)$ is given as 
\begin{equation}
\pw(\ell) = \sum_{n=0}^{\ell-1} (1-\kappa-\mu)^{n} \mu (1-\nu)^{\ell-n-1},
\label{p_w}
\end{equation}
where $\nu$ is the probability of W becoming I by single contact with neighboring I.
Finally, $\pr(\ell)$ is the probability that a node is in state R when it contacts $\ell$ neighbors in state R in the absorbing state. Using the relation $\ps(\ell)+\pw(\ell)+\pr(\ell)=1$, one can determine $\pr(\ell)$ in terms of $\ps$ and $\pw$. 

The order parameter $m$ that a randomly chosen node is in state R after the system falls into an absorbing state is given as
\begin{equation}
m=\sum_{q=1}^{\infty}P_{d}(q) \sum_{\ell=1}^{q} \binom{q}{\ell} r^{\ell}(1-r)^{q-\ell} \pr(\ell),
\label{rho_r}
\end{equation}
where $P_{d}(q)$ is the probability that a node has degree $q$ and $r$ is the probability that an arbitrarily chosen edge leads to a node in state R in the absorbing state. Using the local tree approximation, we define $r_n$ similarly to $r$ but now at the tree level $n$. 

The probability $r_{n+1}$ can be derived from $r_n$ as follows:
\begin{widetext}
\begin{equation} \label{eq:sce}
r_{n+1}=\sum_{q=1}^{\infty} \frac{qP_d(q)}{z} \sum_{l=0}^{q-1} \binom{q-1}{\ell} r_n^{\ell}(1-r_n)^{q-1-\ell} \pr(\ell)\equiv f(r_n),
\end{equation}
\end{widetext}
where the factor $qP_d(q)/z$ is the probability that a node connected to a randomly chosen edge has degree $q$. As a particular case, when the network is an ER network having a degree distribution that follows the Poisson distribution, i.e., $P_d(q)=z^q e^{-z}/q!$, where $z=\sum_q qP_d(q)$ is the mean degree, the function $f(r_n)$ is reduced as follows:
\begin{equation} \label{eq:f_q}
f(r_n)=1-\Big(1-\dfrac{\mu}{\kappa + \mu - \nu}\Big)e^{-(\kappa + \mu)z r_n} - \dfrac{\mu}{\kappa + \mu - \nu}e^{-\nu z r_n}.
\end{equation}

Eq.~(\ref{eq:sce}) reduces to a self-consistency equation for $r$ for given reaction rates in the limit $n\to \infty$. Once we obtain the solution of $r$, we can obtain the outbreak size $m$ using Eq.~(\ref{rho_r}). For ER networks, however, $m$ becomes equivalent to $r$ so that the solution of the \self equation Eq.~(\ref{eq:sce}) yields the order parameter. Thus we define $F(m) \equiv f(m)-m $ so that the order parameter satisfies the following equation
\begin{widetext}
\begin{equation} \label{eq:g_m}
F(m)=1-e^{-(\kappa + \mu)z m}\Big(1-\dfrac{\mu}{\kappa + \mu - \nu}\Big) - \dfrac{\mu}{\kappa + \mu - \nu}e^{-\nu z m}-m =0.
\end{equation}
\end{widetext}

When $m$ is small, the function $F(m)$ is expanded as follows:
\begin{equation}
F(m)\approx am + bm^{2}+cm^{3}+O(m^{4}),
\end{equation}
where
\begin{eqnarray}
a&=&(\kappa z -1) \\
b&=& -\dfrac{1}{2}(\kappa ^2 +\kappa \mu - \mu \nu)z^2 \\
c&=&\dfrac{1}{6} \Big ( \kappa ^3 +2\kappa ^2 \mu + \kappa \mu(\mu - \nu)-\mu \nu(\mu+\nu) \Big )z^3
\end{eqnarray}
with the mean degree $z$. $ m=0 $ is a trivial solution of $F(m)=0$. $c$ is supposed to be negative. 

We remark that although non-trivial solutions exist for appropriate values of $a$ and $b$, some of them might be physically irrelevant for a given initial condition. We can find out the relevancy by checking the stability of the fixed points of $r_{n+1}=f(r_{n}) $. If we impose a small perturbation to the steady state solution $r^*$ of Eq.~(\ref{eq:sce}), we obtain the recursive equation as 
\begin{equation}
r^{*}+\delta r_{n+1}\approx f(r^{*})+\frac{df}{dr}\Big|_{r=r^*} \delta r_n,
\end{equation}
which leads to 
\begin{equation}
\eta \equiv \dfrac{\delta r_{n+1}}{\delta r_{n}}=\frac{df}{dr}\Big|_{r=r^*}.
\end{equation}
If $\eta < 1$ ($> 1$), then the steady state solution $r^*$ is stable (unstable). Note that $ \eta<1 $($ >1 $) iff $ F(m^{*})<0$($ >0 $). If $ a<0 $, the trivial solution $ m=0 $ becomes stable so that other non-trivial solutions cannot be accessible. On the other hand, since $F(1)$ is always negative, the condition $a>0 $ guarantees the existence of a non-trivial solution that is stable and physically accessible. This shows $a=0$ at the transition point which implies $\kappa_{c}=1/z$.  

\subsection{The condition of occurring a discontinuous transition}

The solution of $F(m)=0$ within the order of $m^3$ is $m=0$ and $m_*^{\pm}$, where 
\begin{equation}\label{key1}
m_*^{\pm}=-\dfrac{b}{2c}\pm \sqrt{\dfrac{b^{2}}{4c^{2}} -\dfrac{a}{c}}.
\end{equation}
When a discontinuous phase transition occurs at $a=0$, i.e., at $\kappa_c=1/z$, $-b/c$ is still positive. Since $b=0$ at 
\begin{equation}\label{key2}
\kappa_b = \dfrac{-\mu+\sqrt{\mu^2 +4\mu \nu}}{2},
\end{equation}
and $b < 0$ for $\kappa > \kappa_b$, $b$ has to be positive at $\kappa_a$ as long as $\kappa_a < \kappa_b$. This is the condition for a discontinuous transition to occur at $\kappa_c$ with a nonzero order parameter $m_*=-b/c$. The condition is rewritten in another form as 
\begin{equation}\label{key3}
\dfrac{1}{z} < \dfrac{-\mu+\sqrt{\mu^2 +4\mu \nu}}{2}.
\end{equation}

\section{The size and lifetime distributions of critical branching trees}\label{cb_size_lifetime}

\begin{figure}[h]
\includegraphics[width=1.0\linewidth]{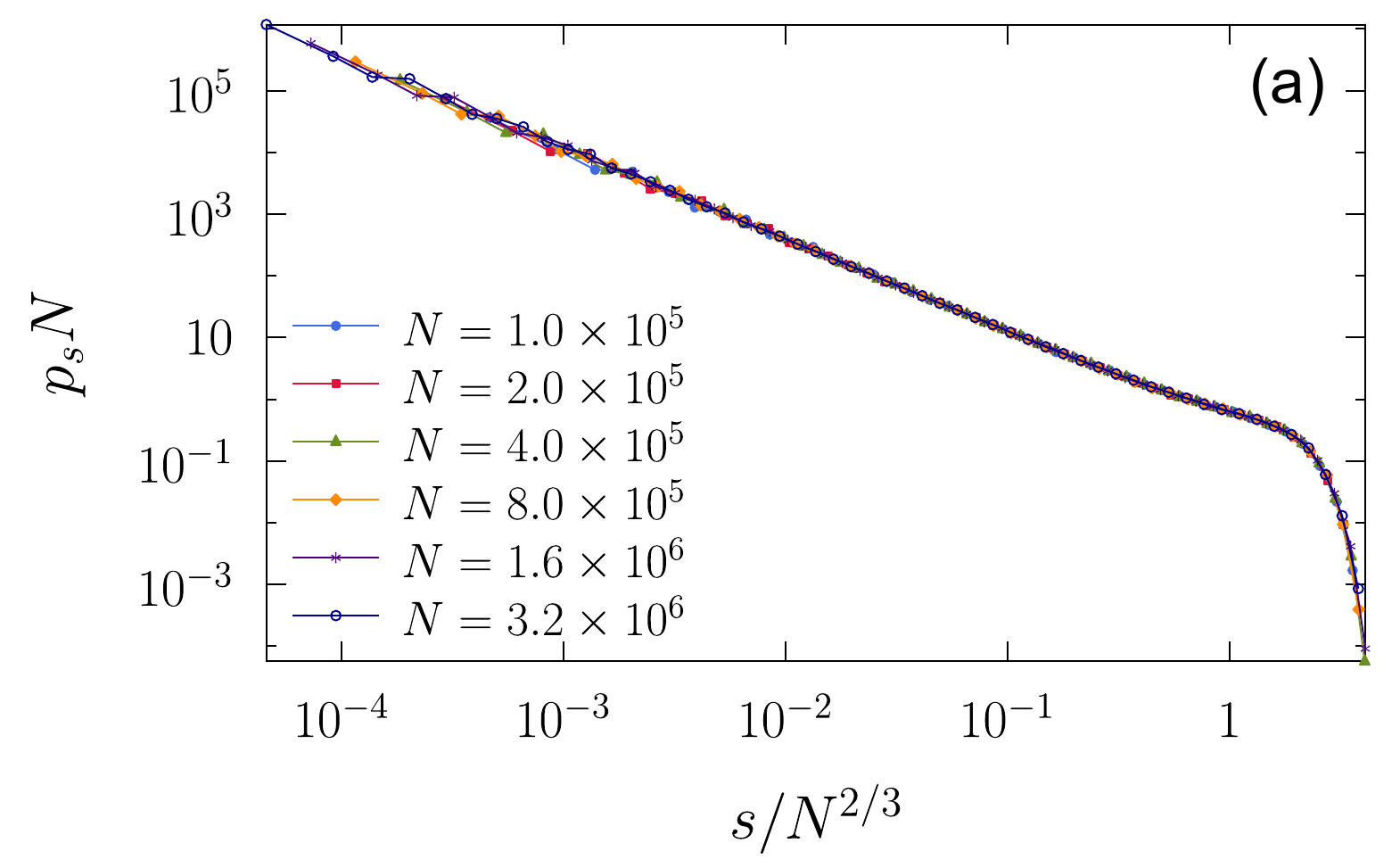}
\includegraphics[width=1.0\linewidth]{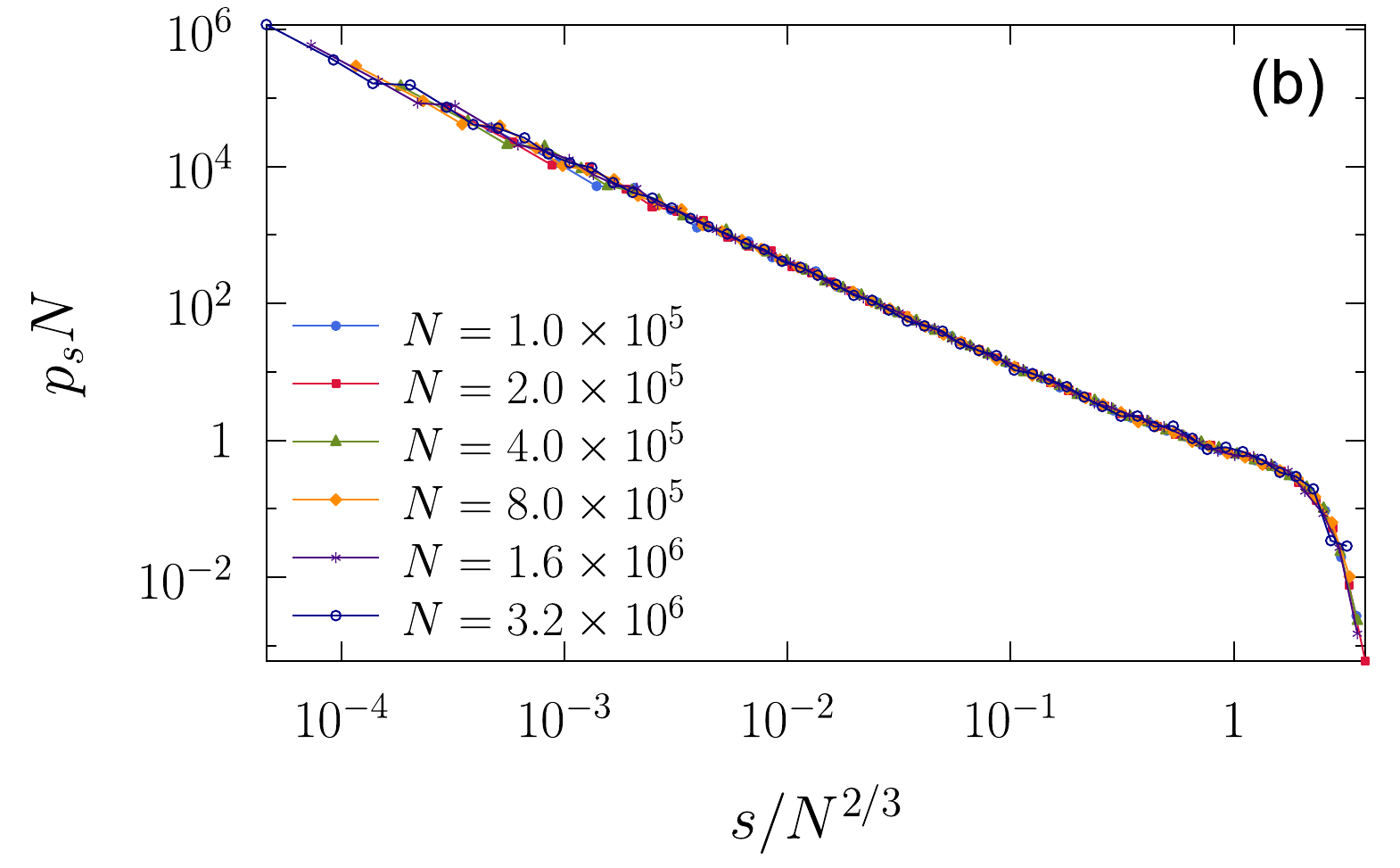}
\caption{Scaling plot of the size distirbution of CB trees on (a) ER networks with mean degree $z=8$ and (b) fully connected networks with degree $z=N-1$ for different system sizes. From a randomly selected seed, a branch is constructed to each neighbor with probability $1/z$. Repeatedly each of the offsprings makes a branch to their neighbors with the same probability $1/z$. We find for both cases that the size distribution of the tree decays a power-law way with the exponent $\tau_s \approx 3/2$ and there exists a characteristic size  $s^*\sim N^{2/3}$ for the CB trees.} 
\label{fig:er_size}
\end{figure}

\begin{figure}[h]
\includegraphics[width=1.0\linewidth]{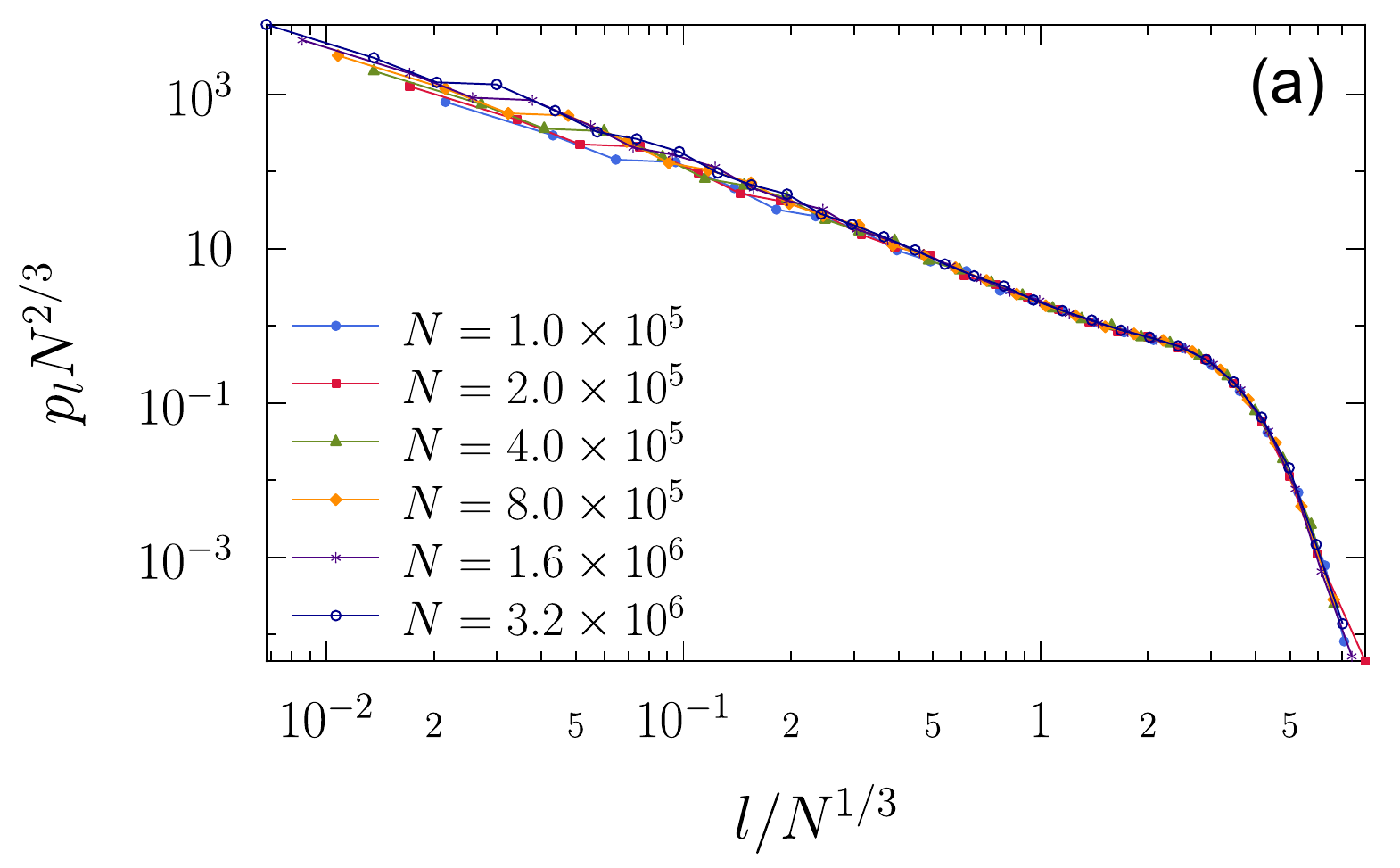}
\includegraphics[width=1.0\linewidth]{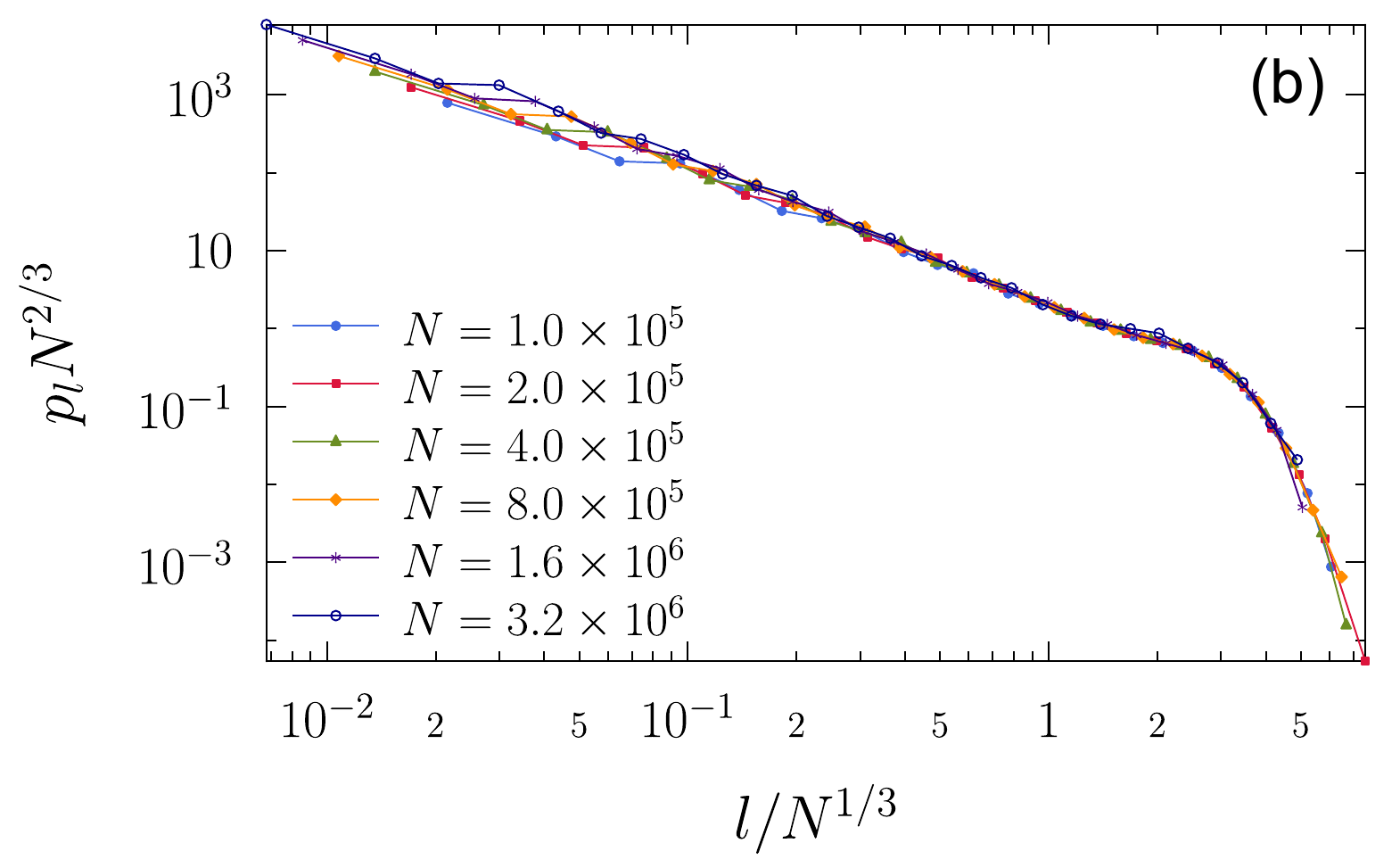}
\caption{Scaling plot of the lifetime distribution of CB trees on (a) ER networks with mean degree $z=8$ and (b) fully connected networks with degree $z=N-1$ for different system sizes. From a randomly selected seed, a branch is constructed to each neighbor with probability $1/z$. Each of the offsprings makes a branch to its neighbors with the same probability $1/z$. This process is repeated successively. We find that the lifetime distribution of the tree decays in a power-law way with the exponent $\tau_{\ell} \approx 2$ and there exists a characteristic size  ${\ell}^*\sim N^{1/3}$ for the CB trees.} 
\label{fig:er_lifetime}
\end{figure}

Here we check if the scaling features of $O(N^{2/3})$ and $O(N^{1/3})$ for the size and the lifetime, respectively, of a CB tree at a transition point $z_c$ are valid even for any $z \gg z_c$. We obtain the distributions of the size and the lifetime of the critical branching trees on ER random networks with mean degree $z=8$ and on fully-connected networks with different system sizes, respectively. Performing numerical simulaitons with different system sizes $N$, we obtain that the size distribution of CB trees follows a power law, $p_s\sim s^{-\tau_s}\exp(-s/s^*)$, where $\tau_s\approx 3/2$ and $s^*\sim N^{2/3}$ (see Fig.\ref{fig:er_size}). The life time distribution also follows a power law $p_{\ell}(\ell)\sim \ell^{-\tau_{\ell}}\exp(-\ell/\ell^*)$ with $\tau_{\ell}\approx 2$ and $\ell^*\sim N^{1/3}$ (see Fig.\ref{fig:er_lifetime}). This numerical results are consistent with the analytic results using the generating function method for a CB tree.  

\begin{figure}[H]
\includegraphics[width=1.0\linewidth]{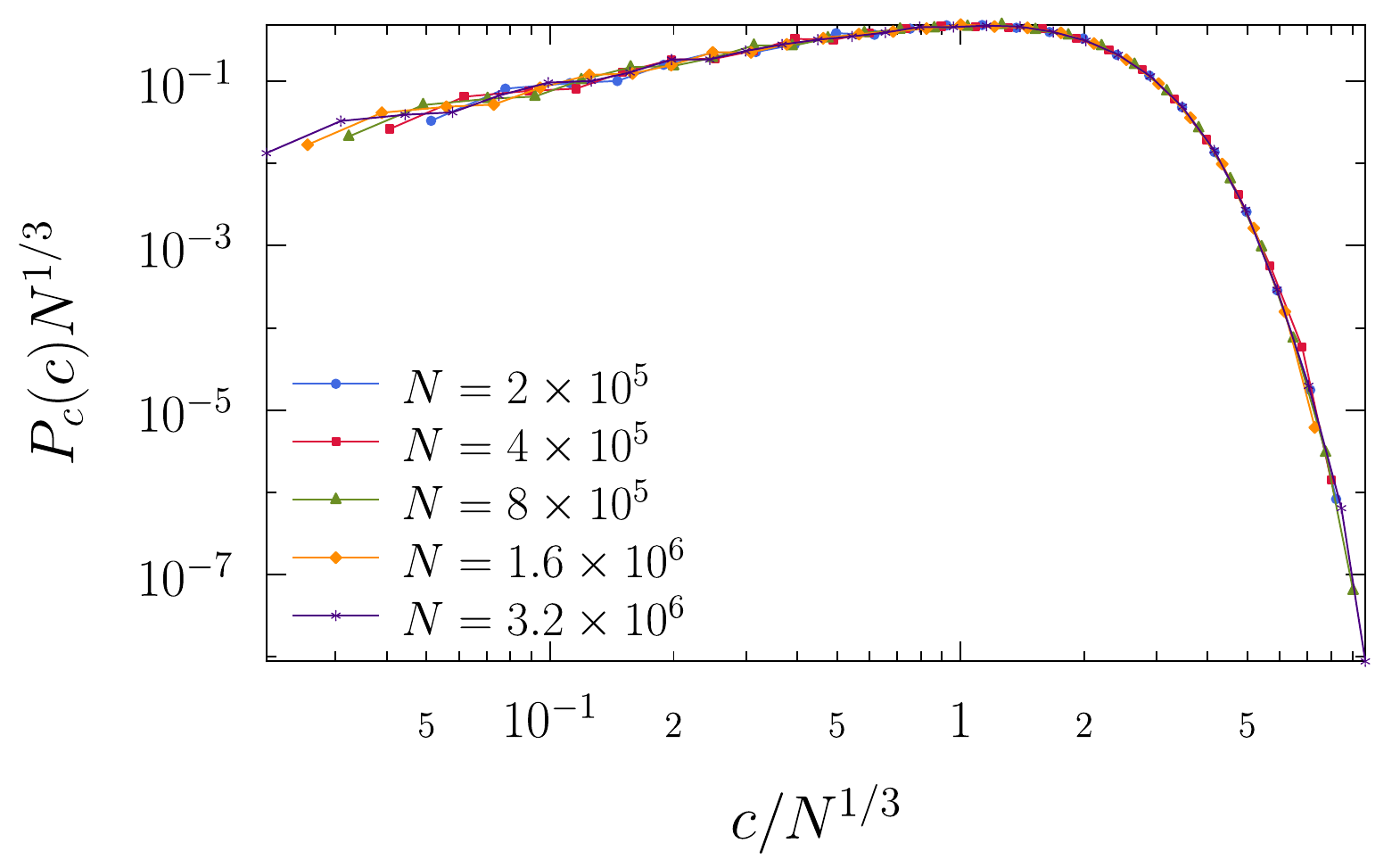}
\caption{Scaling plot of the distribution $P_c(c)$ of the lengths $c$ of loops (i.e., cycle) that are formed by the reaction $W+I\to 2I$ for different system sizes. The scaling is in the form of $P_c(c)N^{1/3}$ vs $c/N^{1/3}$. Here loop length is defined as one plus the sum of the distances from the generations $n_I$ and $n_w$ to their last common ancestor. The nodes $i$ and $j$ are connected by the reaction $W+I\to 2I$ and make a loop. Data are well collapsed onto a single curve and  loop lengths are scaled by a characteristic scale $\sim N^{1/3}$.} 
\label{fig:looplength}
\end{figure}

\section{Loop-length distribution for the reaction $W+I\to 2I$ (Fig.\ref{fig:looplength})}\label{app:loop_length}

\section{How to determine effective degrees of each node}\label{effective_degree}

As defined in the main text, to determine the effective degrees of each node for an infinite avalanche, we needed to count the number of each type of links of a certain node $i$ following which we can access $O(N)$ nodes along only the same type of links. Such links of the node $i$ are called viable links. To implement this counting in finite systems, we present a method at a certain graph with mean degree $z$.

First, we generate an ER graph with mean degree $z$ for each type of links, then determine a GMCC using an efficient algorithm presented in~\cite{mcc_algo}. Second, to determine the effective degree $k_A(i)$ of a node $i$ for A-type links, we consider a network that consists of all nodes in the GMCC and only A-type of links, which is denoted as A-GMCC. 
Next, to determine the viable links of a node $i$ in the A-GMCC, we suppose that the node $i$ is removed, and determine an A-GMCC. If this removal does not break the A-GMCC at all, then all A-type links of the node $i$ are viable links of the node $i$. 
If the removal breaks A-GMCC  into more than one clusters, there may be a unique largest one. Then the links of the node that were connected to the largest cluster are the viable links of type A of the node. One may suppose the case that two or more clusters are of the same size; however, its probability would be negligible. Type B viable links are also determined in the same way. This determination can be efficiently implemented using the algorithms in Refs.~\cite{mcc_algo,hdt_algo}.

The avalanche process is implemented straightforwardly with this definition of viable links. First, we obtain effective degrees of each node in the GMCC, and then identify weakened or susceptible state of each node. To trigger an avalanche, we remove a randomly chosen node, which serves as a seed of the avalanche. At the next time step, we recalculate effective degrees of each neighbor of the seed. If there are neighbors whose one or both types of effective degrees are zero, then we remove them. We repeat these processes with all neighbors of the nodes removed at the previous time step. The repeated process continues until no more node is removed.

\end{document}